\documentclass[twocolumn]{aastex62}
\usepackage[]{amsmath}
\PassOptionsToPackage{hyphens}{url}
\usepackage{hyperref}
\usepackage{soul}
\newcommand{\kperp}{k_\perp}

\newcommand{\Vae}{V_{A}}
\newcommand{\nabperp}{\nabla_\perp}
\newcommand{\vperp}{{\bf v}_\perp}
\newcommand{\zhat}{\hat{\bf z}}
\newcommand{\Bperp}{{\bf B}_\perp}
\newcommand{\de}{d_e}

\newcommand{\be}{\begin{equation}}
\newcommand{\ee}{\end{equation}}
\newcommand{\bea}{\begin{eqnarray}}
\newcommand{\eea}{\end{eqnarray}}

\newcommand{\Eq}[1]{Eq.~(\ref{#1})}

\newcommand{\Eqss}[2]{Eqs.~(\ref{#1})--(\ref{#2})}

\graphicspath{{./}{figures/}}

\received{\today}
\revised{--}
\accepted{--}

\shorttitle{Turbulence in pair plasmas}
\shortauthors{Loureiro et al.}

\begin{document}

\title{Turbulence in Magnetized Pair Plasmas}

\correspondingauthor{Nuno F. Loureiro}
\email{nflour@mit.edu}

\author[0000-0001-9755-6563]{Nuno F. Loureiro}
\affil{Plasma Science and Fusion Center, 
Massachusetts Institute of Technology, 
Cambridge, MA 02139, USA}

\author{Stanislav Boldyrev}
\affiliation{Department of Physics, 
University of Wisconsin-Madison, 
Madison, WI 53706, USA}
\affiliation{Space Science Institute, 
Boulder, Colorado 80301, USA}



\begin{abstract} 
Alfv\'enic-type turbulence in strongly magnetized, low-beta pair plasmas is investigated. 
A coupled set of equations for the evolution of the magnetic and flow potentials are derived, covering both fluid and kinetic scales. 
In the fluid (MHD) range those equations are the same as for electron-ion plasmas, so turbulence at those scales is expected to be of the 
Alfv\'enic nature, exhibiting critical balance, dynamic alignment, and transition to a tearing mediated regime at small scales.
The critical scale at which a transition to a tearing-mediated range occurs is derived, and the spectral slope in that range is predicted to be $k_\perp^{-8/3}$ (or $k_\perp^{-3}$, depending on details of the reconnecting configuration assumed).
At scales below the electron (and positron) skin depth, it is argued that turbulence is dictated by a cascade of the inertial Alfv\'en wave, which we show to result in the magnetic energy spectrum $\propto k_\perp^{-11/3}$. 

\end{abstract}



\section{Introduction} \label{sec:intro}
Plasmas formed of electrons and positrons --- so called pair plasmas --- are thought to be important in a variety of astrophysical contexts, including
 pulsar wind nebulae, astrophysical jets and gamma ray bursts~\citep{woosley_gamma-ray_1993,reynolds_matter_1996,wardle_electronpositron_1998}. 
 The recent detection of a binary neutron star merger~\citep{abbott_gravitational_2017}, where pair plasmas are expected to be generated, presents another compelling and timely motivation for the better understanding of pair plasma physics.
 
The abundance of pair plasmas in astrophysical settings is unfortunately not matched in the laboratory. 
Only very recently has the first pair plasma been created~\citep{sarri_generation_2015},  by means of a laser-driven setup. 
The prospect of ever increasing laser intensities has, furthermore, led to the suggestion that QED cascades might be attainable, seeding the copious generation of pairs~\citep[e.g.][]{vranic_electronpositron_2017}.

Simultaneously, there are ongoing efforts to create the first magnetically confined pair plasma~\citep{sunn_pedersen_plans_2012,stenson_debye_2017}. 
It is hoped that these experiments will shed light on pair-plasma dynamics, thus providing a better understanding of these astrophysical objects. 
In the meantime, they have spurred a series of theoretical investigations~\citep{helander_microstability_2014,helander_gyrokinetic_2016,zocco_slab_2017,mishchenko_gyrokinetic_2018} with noteworthy results highlighting their stability properties.\\

Magnetized pair plasmas naturally occurring in astrophysical contexts may be expected to be in a turbulent state; the understanding of this turbulence, and in particular how it may differ from that found in (much better studied) ``normal'' (electron-ion) plasmas, is critical to make sense of astrophysical observations and dynamics.
To focus the discussion, let us consider weakly collisional plasmas, $\omega \gg \nu$, where $\omega$ is the characteristic frequency of the turbulent fluctuations, and $\nu$ is the collision frequency.
In addition, we will specialize to low plasma beta,
$\beta = 8\pi n T/B^2 \ll 1$, where $n=n^+ + n^-$, with $n^+=n^-$ the positron and electron densities, respectively. 
The electron-positron plasma skin depth is then  $d_e=c/\omega_{p}$, where $\omega_{p} = \sqrt{4\pi n e^2/m_e}$ is the plasma frequency (defined with the total particle density). 
Note that in a low beta plasma, $d_e$ is larger than the electron (or positron) Larmor radius (and also than the Debye scale).

At fluid (MHD) scales, $\kperp d_e \ll 1$, where $\kperp$ is the wavenumber of fluctuations perpendicular to the mean magnetic field, the equations governing pair plasma behavior are indistinguishable from the usual MHD equations of ion-electron plasmas (as we will see in Section~\ref{sec:equations}). 
Therefore, at those scales, turbulence will be no different. 

However, this equivalence stops at the kinetic scales, $\kperp d_e\gg1$, due to the absence of a separation between the electron and positron micro scales.
This has qualitative implications for both the properties of the waves that can propagate at those scales, and the physics of magnetic reconnection and the tearing instability, whose role in kinetic-scale turbulence in electron-ion plasmas is now abundantly appreciated~\citep{tenbarge_current_2013,cerri_reconnection_2017,franci_magnetic_2017,loureiro_collisionless_2017,mallet_disruption_2017-1}). 

To the best of our knowledge, there have been so far no attempts to develop a theoretical understanding of pair plasma turbulence at kinetic scales.
Given the weak collisionality of many of the astrophysical environments where pair-plasmas exist, understanding this scale range is key.
Also somewhat surprisingly, direct numerical simulations of kinetic turbulence in pair plasmas are few. \cite{makwana_energy_2015,makwana_dissipation_2017} conducted decaying turbulence runs with a PIC code, focused mostly on recovering MHD-scale behavior, and indeed found good agreement with MHD simulations. 
A set of state-of-the-art PIC simulations of forced pair-plasma turbulence were performed by \cite{zhdankin_kinetic_2017,zhdankin_numerical_2018}.
These again clearly demonstrate agreement with MHD results at fluid scales, in addition to evidence for a kinetic range cascade and particle acceleration efficiency. 
Naturally, even the best simulations feasible on today's computers fall short of accessing true scale separation between the different length scales of the problem (system size, electron skin depth, electron Larmor radius, and resistive and viscous dissipation scales, if in the collisional limit), and so we believe it is accurate to state that pair plasma turbulence remains poorly understood.

In this paper, we derive a set of fluid-like equations which describe low-beta, sub-relativistic pair plasmas dynamics at scales above the Larmor radius (or Debye length, whichever one is largest). 
These equations are then used to derive the expected spectral power laws and break scales of magnetized turbulence in such plasmas.

\section{Equations for a strongly magnetized, low-beta pair plasma.}
\label{sec:equations}
We aim to obtain fluid-like equations for low beta, non-relativistic pair plasma dynamics, covering scales much larger than both the Larmor radius and the Debye scale.

The equation of motion for each fluid species is:
\begin{align}
\label{eq:momentum}
\begin{split}
n^{\pm} m \left(\frac{\partial {\bf v}^{\pm}}{\partial t} + 
{\bf v}^{\pm}\cdot \nabla {\bf v}^{\pm} -
\mu^{\pm} \nabla^2 {\bf v}^{\pm}\right) =\\ 
\pm n^{\pm} e \left({\bf E} + \frac{{\bf v}^{\pm}\times {\bf B}}{c}\right)-
\nabla p^\pm \mp \mathcal R,
\end{split}
\end{align}
where ${\bf v}^{\pm}$ denote the velocity of the positron and electron fluids, respectively, $\mu^\pm$ their kinematic viscosities, $p^\pm$ the pressures, and $\mathcal R$ is the friction between each species. The density $n$ is the same for both species, $n^+=n^- \equiv n/2$. Following~\citet{chacon_fast_2008}, we assume equal temperatures, from which follows that $\mu^+=\mu^- \equiv \mu/2$, and $p^+=p^- \equiv p/2$. Also, we simply take $\mathcal R = (n/2) e \eta {\bf j}$, where $\eta$ is the resistivity.

As usual, we add and subtract both Eqs.~(\ref{eq:momentum}). 
Defining the center-of-mass velocity ${\bf v} = ({\bf v}^+ + {\bf v^-})/2$, and the current ${\bf j} = (n/2) e ({\bf v^+} -{\bf v^-})=c/(4\pi) \nabla\times {\bf B}$, we obtain
\begin{multline}
\label{eq:tot_momentum}
n m\left(\frac{\partial {\bf v}}{\partial t} + 
{\bf v}\cdot\nabla {\bf v} +
\frac{1}{n^2 e^2}{\bf j}\cdot\nabla {\bf j} -
\mu \nabla^2 {\bf v} \right)=\\
\frac{{\bf j}\times{\bf B}}{c} - \nabla p,
\end{multline}
\begin{multline}
\label{eq:current}
\frac{m}{n e^2}\left(\frac{\partial {\bf j}}{\partial t} + {\bf v}\cdot\nabla{\bf j} + {\bf j}\cdot\nabla{\bf v} -\mu\nabla^2{\bf j} \right) = \\{\bf E} +\frac{{\bf v}\times {\bf B}}{c} - \eta {\bf j}.
\end{multline}

We now wish to specialize these equations to the case of a strongly magnetized plasma. 
I.e, we assume that ${\bf B} = B_0\hat z + {\bf B_\perp}$, where $B_\perp/B_0\sim \epsilon$; this ordering then justifies the assumptions $k_\parallel/\kperp \sim v/V_A\sim \omega/\Omega_e\sim\epsilon$, where $k_\parallel$ and $\kperp$ are typical parallel and perpendicular fluctuation wavenumbers, $\Vae=B_0/\sqrt{4\pi\rho}$ is the Alfv\'en velocity based on the guide-field, $\omega$ is the typical frequency of the perturbations, and $\Omega_e$ is the cyclotron frequency.
Conceptually, the procedure is no different from the familiar reduced-MHD orderings~\citep{schekochihin_astrophysical_2009}, with the difference that we will here allow for $\kperp d_e\sim 1$. 

The orderings above imply that the flow is incompressible, $\nabla\cdot {\bf v} = 0$. 
It is therefore convenient to introduce flow and magnetic potentials, defined as
\be
\vperp = \zhat \times\nabperp \phi; \qquad 
\frac{\Bperp}{\sqrt{4\pi\rho}} = \zhat\times\nabperp\psi,
\ee
where $\rho\equiv n m$ is the total mass density.

To obtain an evolution equation for $\phi$, we extract the perpendicular part of the curl of \Eq{eq:tot_momentum};  for $\psi$,  we take the $z$-component of ~\Eq{eq:current} and use $E_z = (1/c)(\partial \psi){\partial t} - \partial \varphi/\partial z$, with $\phi = (c/B_0)\varphi$.
We then arrive at:
\begin{multline}
\label{eq:vort}
\frac{\partial }{\partial t}\nabperp^2\phi + \{\phi,\nabperp^2\phi\} = \\\{\psi,\nabperp^2\psi\}+\Vae\frac{\partial }{\partial z}\nabperp^2\psi + \mu\nabperp^2\phi,
\end{multline}
\begin{multline}
\label{eq:psi}
\frac{\partial}{\partial t}\left(1-d_e^2\nabperp^2\right)\psi + \left\{\phi,(1-d_e^2\nabperp^2)\psi\right\} =\\
\Vae\frac{\partial \phi}{\partial z}  +\eta\nabperp^2\psi -\mu d_e^2\nabperp^4\psi,
\end{multline}
where, in the last equation, $\eta$ has been redefined to now represent magnetic diffusivity, $\eta c^2/4\pi\rightarrow\eta$.

In the limit $d_e\rightarrow 0$ these equations become equivalent to the reduced-MHD equations~\citep{kadomtsev_nonlinear_1974,strauss_nonlinear_1976,schekochihin_astrophysical_2009}, as expected, differing only in the definition of the Alfv\'en speed.

\subsection{Invariants}
In the limit of no dissipation, the above equations yield the following energy invariant:
\be
\label{eq:energy}
\mathcal{E} = \frac{1}{2}\int dV\left\{ \left(\nabla_\perp \psi\right)^2 + d_e^2 \left(\nabla_\perp^2\psi\right)^2 +\left(\nabla_\perp \phi\right)^2\right\}.
\ee
The terms in this integral are, respectively, magnetic energy, parallel kinetic energy and perpendicular kinetic energy from the $E \times B$ motion. A second exact invariant is the generalized cross-helicity:
\be
\mathcal H^C = \int dV \left\{\nabperp^2\phi\left(1-d_e^2\nabperp^2\right)\psi\right\}. 
\ee
Note that, unlike the energy, this invariant is not sign-definite.
\subsection{Linear waves}
Linearization of \Eqss{eq:vort}{eq:psi} straightforwardly yields the dispersion relation for the inertial Alfv\'en wave:
\be
\label{eq:wave}
\omega_l = \pm \frac{k_z \Vae}{\sqrt{1+\kperp^2d_e^2}}.
\ee
Evidently, this is simply a shear Alfv\'en wave in the MHD limit, $\kperp d_e\ll 1$. 
In the opposite (kinetic) case of $\kperp d_e\gg 1$ it becomes $\omega_l\approx \pm k_z \Vae / (\kperp d_e)=\pm\Omega_ek_z/k_\perp$.
Note that this is the {\it only} wave that is supported by~\Eqss{eq:vort}{eq:psi}: the absence of the Hall term in pair plasmas implies that the whistler wave is absent. 
This greatly simplifies the analysis of energy cascades in the kinetic range with respect to electron-ion plasmas (see Section~\ref{sec:kinetic}).

\section{Turbulence in the MHD limit, $\kperp\ll 1/\de$} \label{sec:MHD}
We are now ready to study Alfv\'enic turbulence in pair plasmas. 
As mentioned above, at MHD scales, $\kperp\ll 1/d_e$, the equations are no different from those ruling the behavior of standard ion-electron plasmas. Therefore, we expect all the same results to apply: critical balance~\citep{goldreich_toward_1995}, dynamic alignment~\citep{boldyrev_spectrum_2005,boldyrev_spectrum_2006,mallet_refined_2015} and, following recent work ~\citep{loureiro_role_2017,boldyrev_magnetohydrodynamic_2017,mallet_disruption_2017}, the possible presence of a sub-inertial range  where the energy cascade is mediated by the tearing instability. 

The transition to this qualitatively different regime is expected to occur at the scale $\lambda_c$ at which the growth rate of the tearing mode in a turbulent eddy roughly matches the eddy turn-over-rate at that scale: $\tau_\lambda\sim \lambda^{1/2} L_0^{1/2}/V_{A}$~\citep{boldyrev_spectrum_2006}. 
If the plasma is sufficiently collisional that no kinetic scales are ever important, the tearing instability in the turbulent eddies will be of the MHD type. 
Hence, as in the MHD case~\citep{loureiro_role_2017,mallet_disruption_2017,boldyrev_magnetohydrodynamic_2017}, we predict a transition at
\be
\lambda_c/L_0 \sim S_{L}^{-\alpha},
\ee
where $L_0$ is the outer scale of the turbulence, $S_{L} = L_0 \Vae/\eta$ is the Lundquist number, and $\alpha=4/7$ for a Harris-sheet~\citep{harris_plasma_1962} type magnetic profile, or $\alpha=6/11$ for a sinusoidal magnetic profile.
Below this scale, the prediction is that the spectral slope steepens to $\kperp^{-11/5}$ or $\kperp^{-19/9}$, respectively, for those two cases.

However, if the plasma is weakly collisional, we have to instead allow for collisionless tearing of MHD size eddies~\citep{loureiro_collisionless_2017,mallet_disruption_2017-1}. 
Following~\citet{zocco_slab_2017}, the relevant expressions for the growth rate of a tearing mode with wavenumber $k$ in an eddy whose shortest dimension is $\lambda$ (and where, therefore, $B_\perp$ varies on the scale $\lambda$) are:
\be
\gamma\sim k V_{A,\lambda}(\Delta' d_e)^2 d_e/ \lambda,
\ee
at low values of the tearing parameter $\Delta'(k)$; and
\be
\gamma\sim k V_{A,\lambda} d_e/ \lambda,
\ee
for large $\Delta'(k)$.
In these expressions, $V_{A,\lambda}$ is the Alfv\'en speed determined with $B_\perp$ at scale $\lambda$ (i.e., with the reconnecting field).

For a Harris-type magnetic field profile, where $\Delta' \lambda \sim 1/(k \lambda)$, matching these two expressions yields $k_{max} \lambda \sim d_e/\lambda$, and $\gamma_{max} \sim (V_{A,\lambda}/\lambda) (d_e^2/\lambda^2)$. 
Imposing $\gamma_{max}\tau_\lambda\sim 1$ yields a transition scale to the tearing-mediated range defined as
\be
\label{lambda-crit-harris}
\lambda_{c}/L_0 \sim (d_e/L_0)^{8/9}.
\ee

Following~\citep{boldyrev_magnetohydrodynamic_2017,loureiro_collisionless_2017}, it is not difficult to see that fluctuation spectrum expected in the range $d_e \ll \lambda \ll \lambda_c$ is
\be
\label{eq:harris-spectrum}
E(\kperp)d\kperp \sim \varepsilon^{2/3}d_e^{-4/3}\kperp^{-3}d\kperp.
\ee

For a sinusoidal-like magnetic profile, where $\Delta'\lambda \sim 1/(k \lambda)^2$, the same reasoning leads instead to: 
\bea
\label{lambda-crit-sin}
\lambda_c/L_0 &\sim&(d_e/L_0)^{6/7},\\ E(\kperp)d\kperp &\sim& \varepsilon^{2/3}d_e^{-1}\kperp^{-8/3} d\kperp.
\eea

\section{Turbulence at kinetic scales, $\kperp \de\gg 1$} 
\label{sec:kinetic}
We now wish to address the range of scales intermediate between the skin-depth and the greater of the Larmor radius or the Debye length. 

Two important observations simplify the picture significantly. 
One, already made, is that the only wave accommodated by our equations is the inertial Alfv\'en wave, whose dispersion relation is given by \Eq{eq:wave}.
Unlike in electron-ion plasmas, here there is thus no ambiguity about which wave cascade will be important.
The second is the fact that 
since electron inertia breaks the frozen flux condition, fluctuations in the pair plasma kinetic range are unfrozen, implying that reconnection, and tearing, cannot occur in that range\footnote{Magnetic reconnection, as well as the tearing instability, require frozen-flux to hold everywhere except in a small boundary layer. Without frozen-flux, the field cannot build the tension that is required to reconnect. Note that this is distinct from simple field line diffusion.}. 

In other words, in a pair plasma, any current sheets and flux-ropes that might naturally form due to turbulent dynamics must be confined to the fluid range. 
Direct numerical support for this observation is yielded by the statistical analysis of current sheets performed by \citet{makwana_energy_2015}, who conclude from PIC simulations that most current sheets indeed exhibit a thickness on the order of $d_e$. 
A related observation is reported in the numerical simulations of ~\citet{loureiro_fast_2013}, where tearing-mode-generated magnetic islands are shown to exhibit a minimum saturation amplitude of $d_e$. 

Let us thus derive the power spectrum of the inertial Alfv\'en fluctuations assuming the cascade of energy from large to small scales. Afterwards, we will discuss the consequences of the conservation of cross-helicity.
The first pertinent observation is that, for $\kperp d_e\gg1$, the first term in the energy invariant \eqref{eq:energy} becomes negligible compared to the second. 
We are thus led to expect rough equipartition between the second and the third terms. Therefore, at these scales, the energy flux is expected to scale as
\be
\label{eq:energy_flux}
{\kperp^2\phi_\lambda^2}/{\tau_\lambda}\sim \varepsilon,
\ee
where $\tau_\lambda$ is the eddy turn-over-time at scale $\lambda\sim\kperp^{-1}$, which we estimate as
\be
\tau_\lambda = 1/\omega_{nl}\sim 1/( k_\perp^2\phi_\lambda),
\ee
from balancing the two terms on the left-hand-side of \Eq{eq:vort}.
Combining these two results yields
$\phi_\lambda \sim \varepsilon^{1/3}\kperp^{-4/3}$. 
The spectrum of $\phi_\lambda^2$ should thus scale as 
\be
E_\phi(\kperp)d\kperp\sim \varepsilon^{2/3}\kperp^{-11/3} d\kperp.
\ee

Equipartition implies that $\phi_\lambda^2 \sim d_e^2 B_\lambda^2$, indicating that the magnetic energy spectrum should similarly scale as $\kperp^{-11/3}$.

Finally, we will postulate that the fluctuations at these scales are critically balanced, $\omega_l\sim\omega_{nl}$. 
At these scales, the linear wave, \Eq{eq:wave}, becomes $\omega_{l}\sim k_\parallel \Vae / (\kperp d_e)$.
We then obtain:
\be
k_\parallel \sim \varepsilon^{1/3}d_e\Vae^{-1/3}\kperp^{5/3}.
\ee

It is interesting to point out that similar scalings for the spectrum and the anisotropy have been previously obtained for the whistler waves and the inertial kinetic Alfv\'en waves in the interval $k_\perp d_e\gg 1$
~\citep[e.g.][]{biskamp_two-dimensional_1996,biskamp_electron_1999,meyrand_universal_2010,andres_two-fluid_2014,chen_nature_2017}.
Although those systems are physically different, the mathematical structure of their governing equations in the inertial regime is identical to that resulting from our system (\ref{eq:vort})~\&~(\ref{eq:psi}).

We now examine what condition is imposed on the turbulent fluctuations by the conservation of  cross-helicity. 
This quantity also cascades toward small scales; however, because it is not sign-definite, one might legitimately expect some cancellation in the integrand. 
We therefore propose that
 the cross-helicity flux should be estimated as 
\begin{eqnarray}
\left(\kperp^2\phi_\lambda\right)\left(d_e^2\kperp^2\psi_\lambda\right)R_\lambda/\tau_\lambda \sim\varepsilon^c,
\end{eqnarray}
where $R_\lambda$ is the dimensionless cancellation factor at scale $\lambda$.  
In the inertial range we estimate  $\kperp^2d_e^2\psi_\lambda^2\sim\phi_\lambda^2$ from the energy invariant, so the cross-helicity flux can be written as $\kperp d_e (\kperp^2\phi_\lambda^2)\theta_\lambda/\tau_\lambda\sim\varepsilon^c$. 
This expression is consistent with the constant energy flux, \Eq{eq:energy_flux}, only if 
\be
R_\lambda\propto 1/(\kperp d_e).
\ee
We therefore predict that corss-helicity cancellation is scale-dependent  and it progressively increases with the wavenumber.

\section{Discussion and conclusions} \label{sec:conclusions}
In this paper, we have derived a set of fluid equations to describe turbulence in low beta, strongly magnetized, pair plasmas. 
Both the MHD ($\kperp d_e\ll 1$) and the kinetic ($\kperp d_e\gg 1$) ranges were examined. 
In the former, the equations are formally the same as the usual equations of reduced-MHD. 
Therefore, turbulence at those scales is expected to be no different, except in the case where the tearing-mediated range is governed by collisionless tearing, whose details in pair plasmas differ from electron-ion plasmas.
At the kinetic scales, it is observed that reconnection cannot take place. Therefore, the only mechanism available to determine the energy cascade is the interaction of inertial Alfv\'en waves. The energy spectrum that is thus produced is computed.

In summary, the magnetic energy spectrum is therefore expected to scale as $\kperp^{-3/2}$ up until the beginning of the tearing-mediated range, $\sim \lambda_c^{-1}<d_e^{-1}$, whereupon it becomes $\kperp^{-8/3} - \kperp^{-3}$ (depending on whether the magnetic profile in the tearing-unstable eddies is best described by a Harris or a sinusoidal profile). 
At the kinetic scales, $\kperp d_e\gg 1$, the expectation is that the slope becomes $\kperp^{-11/3}$.

Our equations, being fluid in nature, ignore potentially important kinetic effects such as Landau damping. 
However, for the inertial Alfv\'en wave, the resonance condition $\omega_l/k_z = V_A/(\kperp d_e)\sim v_{th,e}$ indicates that Landau damping may only become significant at the Larmor radius scale, $\kperp \rho_e\sim 1$, i.e., at length scales smaller than the range of validity of our study.

In numerical simulations, where necessarily the scale separation between the system size and the skin-depth is never sufficiently large, the distinction between $\lambda_c$ and $\de$ may not be enough to warrant observation of the tearing-mediated range.
However, recent PIC simulations of turbulence driven by the magnetorotational instability in a pair plasma~\citep{inchingolo_fully_2018} exhibit a $\kperp^{-3}$ magnetic energy spectrum at scales $\kperp d_e<1$, which we are tempted to suggest as evidence of the tearing-mediated range, Eq. \eqref{eq:harris-spectrum}. 
MRI-driven turbulence is a more complex situation than we envisage here and, therefore, caution is required in such extrapolation.
However, there is some numerical evidence that it may share some of the attributes that we associate with Alfv\'enic turbulence~\citep{kunz_magnetorotational_2016,zhdankin_universal_2017,walker_nature_2016}, which we think lends credence to our suggestion.

As we emphasize in the Introduction to our paper, pair plasma dynamics is of great interest in a variety of astrophysical scenarios, and it is expected that a better understanding of its turbulent behavior and spectrum of turbulent fluctuations may help to elucidate important aspects such as particle acceleration efficiency~\citep{zhdankin_kinetic_2017}.
But, in addition, it is worth remarking a less direct, but important and useful, aspect of investigating pair plasmas: the fact that they can sometimes offer interesting insights into how ``regular'' electron-ion plasmas behave (an example is magnetic reconnection~\citep{bessho_fast_2007,chacon_fast_2008}). 
Similarly here, in addition to their intrinsic interest, we hope that these results may shed light on how kinetic turbulence in regular plasmas might work; for example, the reasoning behind the tearing-mediated range is qualitatively the same as in electron-ion plasmas, with the advantage that pair plasmas can be computationally simulated much more efficiently.

\acknowledgments
Discussions with E.~V. Stenson and D.~A. Uzdensky are gratefully acknowledged.
NFL was supported by
the NSF-DOE Partnership in Basic Plasma Science and
Engineering award no. DE-SC0016215, and by the NSF
CAREER award no. 1654168. SB was partly supported by the
NSF grant PHY-1707272, NASA grant 80NSSC18K0646,  
and by the Vilas Associates Award from the University
of Wisconsin - Madison.

\end{document}